%% file: paper.tex
\documentclass[pss]{wiley2sp} 

\usepackage{graphicx}
\usepackage{dcolumn}
\usepackage{amsmath}
\usepackage{amsfonts}
\usepackage{amssymb}
\usepackage{bbm}
\usepackage{verbatim}
\usepackage{epstopdf}

\usepackage{bm}

\newcommand{\braket}[1]{\langle #1 \rangle}

\newcommand{\R}{\mathrm{R}}

\newcommand{\e}{\varepsilon}

\renewcommand{\P}{\boldsymbol{P}}

\newcommand{\T}{\mathcal{T}}
\newcommand{\W}{\boldsymbol{\mathcal{W}}}

\newcommand{\hl}[1]{ \textcolor{blue}{#1} }
\newcommand{\HL}[1]{ \textcolor{red}{#1}}

\renewcommand{\hl}[1]{{#1}}
\renewcommand{\HL}[1]{{#1}}

\begin{document}

\title{Heat, molecular vibrations, and adiabatic driving in non-equilibrium transport through interacting quantum dots}

\titlerunning{Heat transport in quantum dots}

\author{%
  F. Haupt\textsuperscript{\textsf{\bfseries 1,2}},
  M. Leijnse\textsuperscript{\textsf{\bfseries 3,4}},
  H.~L. Calvo\textsuperscript{\textsf{\bfseries 1,2,5}},
  L. Classen\textsuperscript{\textsf{\bfseries 1,2}},
  J. Splettstoesser\textsuperscript{\textsf{\bfseries 1,2}}
  M.~R. Wegewijs\textsuperscript{\Ast,\textsf{\bfseries 1,2,5}}}
\authorrunning{F. Haupt et al.}

\mail{e-mail
  \textsf{m.r.wegewijs@fz-juelich.de}}
  
\institute{%
  \textsuperscript{1}\,Institute for Theory of Statistical Physics, RWTH Aachen, 52056 Aachen, Germany\\
  \textsuperscript{2}\,JARA- Fundamentals of Future Information Technology \\
  \textsuperscript{3}\,Center for Quantum Devices, Niels Bohr Institute, University of Copenhagen, DK-2100 Copenhagen, Denmark\\
  \textsuperscript{4}\,Solid State Physics and Nanometer Structure Consortium (nmC@LU), Lund University, 221 00 Lund, Sweden\\
  \textsuperscript{5}\,Peter Gr{\"u}nberg Institut, Forschungszentrum J{\"u}lich, 52428 J{\"u}lich,  Germany}

\received{XXXX, revised XXXX, accepted XXXX} 
\published{XXXX} 


\abstract{
\abstcol{In this article we review \hl{aspects} of charge and heat transport in interacting quantum dots and molecular junctions \hl{under stationary and time-dependent non-equilibrium conditions due to finite electrical and thermal bias.} In particular, we discuss how \hl{a discrete level spectrum} can be beneficial for thermoelectric applications, and investigate the detrimental effects of molecular vibrations on the efficiency of a molecular quantum dot as an energy converter.
}
{
In addition, we consider the effects of a slow time-dependent modulation \hl{of applied voltages} on the transport properties of a quantum dot and show how this can be used as a spectroscopic tool \hl{complementary to standard \HL{dc-measurements}.} \hl{Finally, we combine time-dependent driving with thermoelectrics in a double-quantum dot system} -- a nanoscale analogue of a cyclic heat engine -- \hl{and} discuss its operation and the main limitations to its performance.}}



\maketitle   

\input{include-intro-FH.tex}

\input{include-method.tex}
\input{include-heat.tex}
\input{include-intro-pumping.tex}
\input{include-pumping-js.tex}
\input{include-heat-pumping.tex}

\begin{acknowledgement}
We acknowledge H. Schoeller for initial collaboration within the SPP-1243, K. Flensberg for collaborations, and financial support from the 
DFG under Contract No. SPP-1243, the Excellence Initiative of the German Federal State Government, and from the Ministry of Innovation NRW.
ML additionally acknowledges financial support from the Swedish Research Council (VR) and nmC@LU.
\end{acknowledgement}


\providecommand{\WileyBibTextsc}{}
\let\textsc\WileyBibTextsc
\providecommand{\othercit}{}
\providecommand{\jr}[1]{#1}
\providecommand{\etal}{~et~al.}

\end{document}

%% file: include-intro-FH.tex

\section{Introduction\label{intro}}
Quantum dots are small man-made structures, with a length scale typically ranging from nanometres to a few microns.
They can be realized in semiconductor heterostructures~\cite{Johnson92,Hanson07rev}, semiconductor nanowires~\cite{DeFranceschi03,Jespersen06,Nilsson09}, nanotubes~\cite{Bockrath97,Tans97}, graphene~\cite{Guttinger12}, or 
even with single molecules~\cite{Park00,Kubatkin03,vanderZant06}.  The name refers to the zero-dimensional nature of \hl{this class of systems}, meaning that the wave function of charge carriers residing on the dot is confined in all spatial directions~\cite{Kouwenhoven01rev}.  
In a transport setup, a quantum dot is tunnel coupled to source and drain electrodes and a dc-current is established by applying a finite source-drain bias voltage. Because of the strong Coulomb repulsion between electrons localized on the dot \hl{and the relatively weak coupling to the electrodes}, charge carriers are transferred through the device one-by-one, and transport can even be completely blocked at low voltages. This phenomenon is known as Coulomb blockade~\cite{Averin86}. In addition to the source and drain, there is often a gate electrode,  which is only capacitively coupled to the quantum dot and can be used to electrostatically control the number of electrons on the dot, as well as the energy cost of adding more electrons, and thereby the conductance of the junction. In such a transistor geometry, the electronic current provides direct information on the \hl{quantum} level-structure and \hl{classical} charging energies of the dot itself. \hl{This generic description provides a good starting point even for systems as small as a single molecule or atom in a junction.}

Applications of quantum dots as ultra-sensitive detectors~\cite{Clerk10}, thermometers~\cite{Giazotto06} or basic building blocks in quantum computers~\cite{Loss98} have already been considered for a long time. Lately, the possibility of exploiting 
their properties for energy conversion or cooling applications have also started to attract considerable attention.  As an example, a refrigeration scheme based on resonant tunneling through quantum dots~\cite{Edwards93} has been recently demonstrated experimentally~\cite{Prance09}.  Theoretically, it has been shown that if a quantum dot is weakly tunnel coupled to a hot and a cold electrode, it can act as an efficient \HL{thermoelectric} energy converter~\cite{Murphy08,Esposito09}.

In conductors, heat is carried dominantly by electrons and the phononic contribution to the heat flow can be often neglected in first approximation. However, phononic heat currents set an ultimate limit to the properties of thermoelectric devices. Minimizing them is crucial to achieve efficient energy conversion and for cooling applications. In molecular quantum dots, phonons (quantized molecular vibrations), 
strongly affect not only the thermoelectric but also the transport properties of the device. In fact, an electric current can excite
vibrations and even drive them out of equilibrium. This in turn affects the conductance of the junction, as well as its stability. For electronic applications, it can therefore be desirable to
design (molecular) quantum dots with good thermal contacts to the
electrodes, or other heat reservoirs.

The study of heat currents is relevant also in the case of time-dependently driven systems, as it carries information on the dissipation associated with the driving~\cite{Moskalets02}. 
Time-dependently driven quantum dots have attracted a lot of attention in recent years, for the possibility of realizing quantized charge pumping~\cite{Kouwenhoven91,Fletcher03,Kaestner08} and reliable single-electron sources~\cite{Feve07}.
These have a realm of applications, ranging from metrology~\cite{Pekola_rev}, to solid-state based quantum information. Spin-pumping has also been demonstrated~\cite{Watson03}. 
\HL{Without thermal driving,} time-dependent driving can be employed as a spectroscopic tool to investigate the dot.
In fact, it has been shown that the charge pumped through the system in response \HL{to} the ac-driving  is sensitive to the microscopic detail of the potential landscape of the dot~\cite{Fletcher03} and, in appropriate modulation setups, it provides information on the coupling asymmetry~\cite{Reckermann10a}, or even on subtle renormalization effects due to tunneling~\cite{Splettstoesser06}.  Furthermore, an ac-driven quantum-dot system can also be operated as a nanoscale {\em cyclic} engine, exchanging heat and work with reservoirs with different temperatures or chemical potentials. 

In this review, we discuss recent progress in the theoretical understanding of several aspects of charge and heat flow in 
quantum dot systems, both with \hl{non-equilibrium conditions} induced by a stationary voltage or temperature gradient and by time-dependent driving of externally applied fields. The paper is organized as follows. 

Section~\ref{sec:method} summarizes the master equation approach employed for calculating the charge, the heat and the spin current through a quantum-dot device, \hl{based on lowest order tunneling processes and an expansion in the driving frequency. }

Section~\ref{sec:heat} focuses on the thermoelectric effect. The physical origin of the thermoelectric effect is explained and 
the desirable electronic properties of a good thermoelectric material are discussed, along with the problems involved in 
finding real systems exhibiting these properties. 
In addition, the different contributions to the phononic heat current are discussed in some detail. All these aspects are then brought together 
in a simple model of a molecular thermoelectric device, where the sharp molecular orbitals provide a large thermoelectric effect, while 
electron--phonon coupling on the molecule, as well as coupling between quantized molecular vibrations and substrate phonon modes in the electrodes, 
\HL{give} rise to significant losses.

Section~\ref{sec:pumping} discusses time-dependent electronic transport
and reviews recent progress on pumping in strongly interacting dots.
Focusing on the slow driving regime, we exploit the geometric nature of
the pumped currents and write them as the flux generated by a
pseudo-magnetic field. As an application example, we consider a
single-level quantum dot driven by time-dependent gate and bias
voltages, where additional non-equilibrium effects are induced by a
finite static bias. We use the pseudo-magnetic field to describe
interaction-induced pumping and to investigate internal properties of
the dot such as spin degeneracy and junction asymmetry in different
regimes of the voltage bias.

Finally, as a further application example, section~\ref{sec:heatpumping}
investigates the heat current in the presence of time-dependent
driving. For the case of a driven double-dot, we show that there are
regimes where not only the charge \HL{current} but also the heat current is
quantized, and that in these regimes \HL{the} double dot can be regarded as a
nanoscale analog of the Carnot engine.

%% file: include-method.tex
 


\section{Microscopic model and transport theory}
\label{sec:method} 
The (molecular) quantum dot setup that we  want to describe can be represented in
general as a central device region that is tunnel coupled to
electronic or bosonic reservoirs, which can usually be considered non-interacting. 
The corresponding Hamiltonian can be written as $H=H_{\rm D}+H_{\rm
res}+H_{\rm coupl}$, where $H_{\rm D}$ is the Hamiltonian of the device
and is typically governed by many-body interactions. Without loss of
generality, we can write $H_{\rm D}=\sum_{\xi}E_{\xi}|\xi\rangle\langle
\xi|$, where $|\xi\rangle$ denote the many-body eigenstates of the
device, with energy $E_{\xi}$.
\hl{The reservoirs are described by a quadratic (non-interacting) Hamiltonian $H_{\rm res}$
and we assume them to be held at separate equilibria with temperatures $T_{r}$ (and chemical potentials} $\mu_{r}$, for the fermionic reservoirs). The
coupling term $H_{\rm coupl}$ is bi-linear in the operators of the
device and the reservoirs.  The coupling to the
reservoirs is quantitatively characterized by the coupling strength
$\Gamma_{r}=2\pi\nu_{r}|t_{r}|^2$, where $t_{r}$ is the single-particle
tunneling matrix element between the device and reservoir $r$, and
$\nu_{r}$ is the density of states of the latter ($\hbar=k_{\rm
B}=e=1$). Both $t_{r}$ and $\nu_{r}$ are often assumed to be energy
independent (wide-band limit). In general,  
the Hamiltonian of the system $H$ can depend on time due to some
external time-dependent driving, i.e. $H(t)=H(\bm{u}(t))$, where
$\bm{u}(t)\equiv\{u_{i}(t)\}$ are the driving parameters. 

\subsection{Master equation approach}
The explicit results presented in the following sections are
based on a master equation approach (unless mentioned otherwise), where the state of the
device is described in terms of the occupation probabilities $P_{\xi}(t)$ of its eigenstates. 
To lowest order in the tunneling, their evolution is governed by the \HL{Markovian} master equation
\begin{equation}
\frac{d}{d t}\P(t)=\W_{t} \P(t), \label{eq:mastereq}
\end{equation}
with $\P(t)=\{P_{\xi}(t)\}$. The kernel $\W_{t}$ 
can be written as a sum of independent contributions from the
reservoirs ${\W_{t}}=\sum_r \W_{r,t}$, where the matrix element
$[\W_{r,t}]_{\xi\xi'}\propto\Gamma_{r}$ represents the probability per unit time that a
tunneling event from/to reservoir $r$ induces the transition
${|\xi'\rangle\to|\xi\rangle}$, 
as given by Fermi's golden rule. The subscript $t$ indicates that these rates have to be evaluated with the parameters of the Hamiltonian ``frozen'' at time $t$.

If the Hamiltonian $H$ is time independent, the evolution kernel is also time independent $\W_{t}\to\W$, and in the long-time limit the system reaches the stationary state $\P^{(0)}$, which satisfies $\W\P^{(0)}=0$. 

Describing the dynamics of a driven systems in terms of Eq.\eqref{eq:mastereq}, we implicitly assumed the time scale of the driving to be much slower than the electron life-time in the system. In this case,  Eq.~\eqref{eq:mastereq} can be solved by means of an 
adiabatic expansion for the occupation probabilities $\P(t)\to \sum_{k\ge0}\P^{(k)}_t$, where $\P^{(k)}_{t}\propto(\Omega/\Gamma)^{k}$ in the case of harmonic driving~\cite{Splettstoesser06}. Here, $\Omega$ is the driving frequency and $\Gamma$ the characteristic tunnel rate of the system. The adiabatic expansion gives rise to the hierarchy of equations~\cite{Splettstoesser06,Cavaliere09}
\begin{equation}\label{eq:hierarchy}
{\rm a)}\ { \W}_t\P^{(0)}_t=0,\qquad {\rm b)}\ { \W}_t\P^{(k)}_t=\frac{d}{dt}\P^{(k-1)}_t.
\end{equation}
Here, $\P^{(0)}_{t}$ is the solution of the problem with all parameters frozen at time $t$. It represents the steady state the system would relax into if it could instantaneously follow the modulation of the time-dependent parameters. We will therefore refer to it as the {\em instantaneous} solution. 
Corrections due to retardation effects are encoded in $\P^{(k>0)}_t$ and are governed by a competition of the time scales of the driving and of the response time contained in $\W_{t}$. The normalization conditions are ${\rm Tr}\{\P^{(0)}_{t}\}=1$, and ${\rm Tr}\{\P^{(k>0)}_{t}\}=0$, where the trace of a vector is defined
as the sum of its components.

The above master equation approach can be rigorously derived~\cite{Splettstoesser06,Cavaliere09}
 starting from a real-time diagrammatic expansion of the reduced density matrix of the system~\cite{Koenig96a,Koenig96b}, and it is valid in the regime of weak coupling to the leads $\Gamma\ll T={\rm min}\{T_{r}\}$ and  slow driving $\Omega\ll\Gamma$. 

\subsection{Charge, spin and heat currents}
We are interested in the charge, heat and spin currents \HL{flowing 
in response to}
a time-dependent modulation of the
parameters of the Hamiltonian and/or to an external bias voltage or
temperature gradient. The electric current flowing out of \hl{lead $r=L,R$} is
found from the derivative of the electron number
$I_{r}(t)=\frac{d}{dt}\big\langle N_{r}\big\rangle$, where
$N_{r}=\sum_{k\sigma}c_{rk\sigma }^{\dag}c_{rk\sigma }$ is the
occupation number operator in lead $r=L,R$. Similarly, the heat and spin \HL{currents are defined}
as $Q_{r}(t)=-\frac{d}{dt}\big\langle
H_{r}-\mu_{r}N_{r}\big\rangle$ and $J_{r}(t)=-\frac{d}{dt}\big\langle
S_{r}^{z}\big\rangle$, respectively, where
$H_{r}=\sum_{k\sigma}\epsilon_{rk\sigma}c_{rk\sigma }^{\dag}c_{rk\sigma }$ is the Hamiltonian of the lead and 
$S_{r}^{z}=\sum_{k\sigma}\frac{\sigma}2c_{rk\sigma }^{\dag}c_{rk\sigma
}$ is the \hl{projection of the lead spin on the quantization axis (chosen along the magnetic field if there is one) and $\sigma=\pm$ stands for spin $\uparrow,\downarrow$.}
A phononic heat current is defined in the same way, but with $\mu_r = 0$ and $H_{r}=\sum_{qr}\omega_{qr}b^{\dag}_{qr}b_{qr}$.

The currents can be evaluated directly from the knowledge of the occupation probability $\P(t)$, being ${R(t)={\rm Tr}\{\boldsymbol{\mathcal{W}}^{R}_t \boldsymbol{P}(t)\}}$,
where $R\in\{I_{r},Q_{r},J_{r}\}$
and the kernels
$\boldsymbol{\mathcal{W}}^{R}_t$ take into account the charge, the heat
and the spin flowing from lead $r$ into the device, \hl{respectively.
For weak coupling to the leads ($\Gamma\ll T$), these kernels read as}
\HL{
\begin{gather}
\big[\boldsymbol{\mathcal{W}}^{I_r}_t \big]_{\xi\xi'} =-e(n_\xi-n_{\xi'})\left[ \boldsymbol{\mathcal{W}}_{r,t} \right]_{\xi\xi'}
,\nonumber\\
\big[\boldsymbol{\mathcal{W}}^{Q_r}_t \big]_{\xi\xi'} = \{E_\xi - E_{\xi'} - \mu_{r}(n_\xi - n_{\xi'})\}\! \left[ \boldsymbol{\mathcal{W}}_{r,t} \right]_{\xi\xi'},\nonumber\\
\big[\boldsymbol{\mathcal{W}}^{J_r}_t \big]_{\xi\xi'}
 =(s_\xi-s_{\xi'})\left[ \boldsymbol{\mathcal{W}}_{r,t}
 \right]_{\xi\xi'}.
\end{gather}
}
Here, $E_\xi$, $n_\xi$ and $s_\xi$ are the energy, the number of
electrons and the spin in the device in the state $|\xi\rangle$,
respectively. 

In the case of slow harmonic driving, the expansion in the driving
frequency carried out for $\P(t)$ results in an analogous expansion for
the current $ R(t) = \sum_{k\ge0}R^{(k)}_t$, where   ${R^{(k)}_{t}=
{\rm Tr}\{\boldsymbol{\mathcal{W}}^{R}_t \boldsymbol{P}^{(k)}_t\}}$ is
the contribution of order $\Omega^{k}$ to the current. The instantaneous contribution $R^{(0)}_{t}$ is the only non vanishing term in the stationary situation, and it is non-zero only if the system is brought out of equilibrium by means of a bias voltage or a temperature gradient. The first order correction $R^{(1)}_{t}$ encodes the effects of adiabatic pumping, and it will be discussed in Sec.~\ref{sec:methods:adiabatic}. Higher order corrections $R^{(k>1)}_{t}$  can in general be neglected for the charge and the spin current ($R\in\{I_{r},J_{r}\}$), as long as $\Omega\ll\Gamma$. This is however not true for  the heat current~\cite{Haupt13}, where contributions to second order in the driving frequency $Q_{r,t}^{(2)}$  account for heating effects due to the ac-driving, see Sec.~\ref{sec:engine}.

%% file: include-heat.tex
\section{Thermoelectric devices and heat currents\label{sec:heat}}
\begin{figure*}[h]
\begin{centering}
  \includegraphics[height=0.2\linewidth]{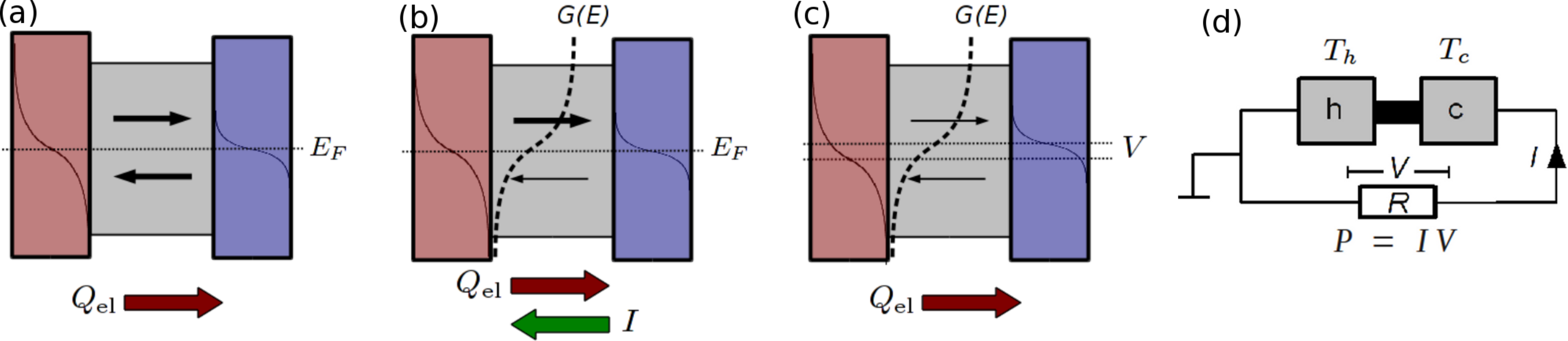}	
	\caption{\label{fig:TEeffect} 
	(a)--(c) Explanation of the thermoelectric effect. Red and blue rectangles indicate the conduction bands of the
	metallic hot and cold electrodes, respectively. The curved line represents the thermal smearing of the 
	Fermi surface around the Fermi energy, $E_F$. 
	(a) If the transport properties of the central region are the same for electrons above and 
	below $E_F$, the temperature difference only results in an (electronic) heat current $Q_\text{el}$, but no 
	net electric current. 
	(b) If, on the other hand, the conductance $G(E)$ of the central region is larger (or smaller) for $E > E_F$, 
	also a net (thermo)electric current $I$ results from the temperature difference. 
	(c) If the thermoelectric current is allowed to flow for some time in an open circuit, a (thermoelectric) voltage $V$ builds up, 
	which eventually cancels the electric current.
	(d) Sketch of thermoelectric circuit converting heat power, $Q$, 
	supplied to the hot electrode, into electric power, $P$, in an electrically driven device (represented by 
	a resistor, $R$).}
\end{centering}
\end{figure*}

In a thermoelectric device, an electric current can be generated as a result of an applied temperature difference $\Delta T$.
The underlying physical mechanism is perhaps easiest to understand in a ballistic device, where 
electrons can be transported between a hot and a cold metallic contact through a central region without loosing energy,
see \HL{Figs.~\ref{fig:TEeffect} (a)-(c)}. 
The larger smearing of the Fermi surface on the hot side results in a net current of high-energy electrons tunneling from the 
hot to the cold electrode, as well as a net current of low-energy electrons tunneling in the opposite direction. 
Therefore, if the conductance is energy independent [Fig.~\ref{fig:TEeffect}(a)], the total electric current is zero and only 
a heat current flows from the hot to the cold electrode.  
If, on the other hand, either high- or low-energy electrons are more easily transported through the central region [Fig.~\ref{fig:TEeffect}(b)],
the electron currents  no longer cancel each other and an electric 
current can flow as a result of the temperature difference,
\hl{in addition to the heat current.} This is the thermoelectric effect. 
Thus, a thermoelectric material must have different transport properties for electrons above and below the Fermi energy.

In an open circuit, the thermoelectric current will lead to charge accumulation in the electrodes and thereby a thermoelectric 
voltage, $V$, which will eventually be large enough to stop the net electric current, see Fig.~\ref{fig:TEeffect}(c). 
The proportionality constant, $S = - V / \Delta T$, is called the Seebeck coefficient or thermopower.
The sign of the voltage depends on the direction of the thermally generated current and therefore on whether the electron or 
the hole conductance is largest. 

\subsection{Thermoelectric efficiency}
In order to perform useful electric work, the setup of Fig.~\ref{fig:TEeffect} has to be modified to include an external electric circuit 
making use of the extracted power, which can simply be considered as a resistor, see Fig.~\ref{fig:TEeffect}(d).
The efficiency $\eta$ of a heat to electric power converter is given by the generated output electric power, $P = IV$, 
divided by the input heat power which has to be supplied to the hot electrode to keep it hot.
If we neglect other heat losses, the needed input heat is equal to the heat current $Q$ flowing from the hot to the cold electrode, and
\begin{align}\label{eq:converter_eta}
	\eta = \frac{I V}{Q}.
\end{align}
Since no heat engine operating between a hot and a cold heat bath can be more efficient than the ideal Carnot process, 
$\eta$ is limited from above by $\eta_C = 1 - T_c / T_h$.

Commonly one considers only the linear regime, where $I = G V + G_T \Delta T$, where $G$ is the electrical conductance and 
$G_T$ the thermal conductance. The Seebeck coefficient is then $S = G_T / G$ and the heat current 
$Q = ( \kappa_\mathrm{el} + \kappa_\mathrm{ph} ) \Delta T$,
where $\kappa_\text{el}$ and $\kappa_\text{ph}$ are the electron and phonon contributions to the thermal conductance, respectively.
A large $\eta$ is related to a large thermoelectric figure of merit, $Z T$, \HL{with $T=(T_{c}+T_{h})/2$}, given by
\begin{align}\label{eq:zt}
	Z T 	= 	\frac{G S^2 T}{\kappa_\text{el} + \kappa_{ph}}.
\end{align}
In the linear response regime, the relation between $Z T$ and $\eta$ is~\cite{Muller08}
\begin{align}\label{eq:zt-efficiency}
	\eta	= 	\frac{\Delta T}{T_h} \frac{\sqrt{1 + ZT} - 1}{\sqrt{1 + ZT} + T_c /T_h }.
\end{align}

Today's best thermoelectrics are heavily doped narrow bandgap semiconductor materials with $ZT \approx 1$.
Inserting $ZT = 1$ in Eq.~(\ref{eq:zt-efficiency}) and assuming $T_h \approx T_c$, one finds $\eta \approx 0.17 \eta_C$. 
Finding bulk materials with high $ZT$ has proven to be more difficult than initially anticipated. One reason is that 
the ratio of the electronic thermal conductance and the electric conductance in bulk materials \HL{follows} the 
Wiedemann-Franz law~\cite{Ashcroft76}
\begin{align}\label{eq:wiedemann-franz}
	\frac{\kappa_\text{el}}{G T} 	= 	\frac{\pi^2 k_B^2}{3 e^2},
\end{align}
and increasing $G$ is therefore accompanied by an increase in $\kappa_\text{el}$. 
Efforts to maximize the efficiency have instead focused on reducing  
$\kappa_\text{ph}$. The challenge is to achieve this without affecting $G$ too much, 
which can be done by introducing effective phonon-scattering centers in the form of heavy-ion species 
with large vibrational amplitudes~\cite{Kim10}.
Recently, nanoscale engineering has been used to achieve this goal, e.g., using superlattice structures with an acoustic mismatch 
between the different layers, resulting in interfaces which scatter phonons more efficiently than electrons, 
which has resulted in impressive figures of merit ($ZT \approx 2.4$ in Ref.~\cite{Venkatasubramanian01}).
Superlattice structures require precise nanoscale engineering and are thus not 
suitable for large-scale production. Fortunately, it seems like the same positive effects can be achieved in 
non-periodic structures, such as nanocomposite materials consisting of a host material filled 
with nanoparticles~\cite{Yang04,Dresselhaus07},  or by introducing roughness 
into Si nanowires~\cite{Hochbaum08}.

Strictly speaking, Eq.~(\ref{eq:wiedemann-franz}) holds only for free electrons, but it is approximately valid 
in most bulk systems. Fortunately, several recent studies, see e.g., Refs.~\cite{Appleyard00,Boese01b,Kubala08}, have shown that the 
Wiedemann-Franz law completely breaks down in nanoscale systems. The reason for this breakdown is either 
electron--electron interaction, or quantum confinement leading to effectively one- or two-dimensional 
structures. It is therefore useful to for the moment forget about the phonons ($\kappa_\text{ph} \rightarrow 0$) and ask the 
question: Which (electronic) material properties would actually maximize $ZT$ as defined in Eq.~(\ref{eq:zt})? And how large would this $ZT$ be?
This problem was studied and solved by Mahan and Soho~\cite{Mahan96}.
They showed that the ideal material is characterized by the conductivity as a function of energy being propotional to the 
$\delta$-function, where the peak should be localized $2.4 k_B T$ above or below the Fermi level (giving an electron-type or 
a hole-type thermoelectric device). 
In this case, $Z T$ goes to infinity and Carnot efficiency can be achieved (meaning reversible thermoelectric operation~\cite{Humphrey02,Humphrey05}). In practice of course  $\kappa_\text{ph} > 0$ and $ZT$ remains finite. Remembering that the origin of the thermoelectric effect is an energy asymmetry in the conductivity [Fig.~\ref{fig:TEeffect}(b)], 
it is rather easy to understand the merits of a $\delta$ function, as it allows perfect energy filtering of the 
electrons passing through the material. 

Mahan and Soho suggested using rare-earth compounds with very sharp electronic $f$-levels to achieve a $\delta$-like
conductivity, but correct alignment of these levels with respect to the Fermi energy is problematic, as is the phonon \HL{contribution} to the heat current. 
A more modern way of introducing a strong energy asymmetry in the conductivity is to take advantage of nanostructuring to reduce 
the effective dimensionality of the system, since the density of states in low-dimensional systems is less smooth.
Initial experiments focused on two-dimensional quantum wells, see e.g., Ref.~\cite{Hicks96}, where the density of states is constant, 
with steps whenever a new 2D band becomes accessible. 
Also one-dimensional systems, such as nanowires~\cite{Boukai08,Hochbaum08} and carbon nanotubes~\cite{Small03,Llaguno04}, have been investigated,
where the density of states has sharp peaks corresponding to the bottom of the 1D bands.
In quantum dots, which are effectively zero-dimensional, the density of states has peaks corresponding to the discrete orbitals, and the 
conductance therefore resembles the $\delta$-function shape.
\HL{Several theoretical and experimental works have investigated the Seebeck
effect in quantum dot devices, see e.g., Refs.~\cite{Beenakker92,Turek02,Sheibner07,Svensson12}.}
However, quantum dots made in inorganic materials have level spacings 
of a few meV at the most, and the density of states therefore looks rather smooth 
at room temperature, which limits the usefulness for most applications. 
We will therefore now turn to molecular devices, since a molecule weakly coupled to electrodes behaves as a quantum dot, 
but with a level spacing which can be much larger than $k_B T$ also at room temperature.

First it should be mentioned, however, that achieving a high $ZT$ is not all that matters for thermoelectric materials. 
Even with $ZT \rightarrow \infty$, Carnot efficiency can only be reached in the limit of infinitely slow, reversible operation, where
the device produces zero output power. It is therefore also interesting to look at intrinsically nonequilibrium quantities such 
as efficiency at maximum output power. Nonetheless, studies have indicated~\cite{Esposito09} that also in such cases a 
$\delta$-like conductivity is desirable.
One problem with a $\delta$-like conductivity is that the actual output power is rather small. For example, if a zero-dimensional 
system is strongly coupled to electrodes to allow for a large current flow, and thereby large output power, the orbital states 
are broadened into Lorentzians, rather than $\delta$ functions, resulting in a decreased efficiency. Therefore, it was the conclusion of 
Ref.~\cite{Nakpathomkun10} that one-dimensional systems might be preferable when a high output power is desired.

\subsection{Molecular thermoelectric devices}
\begin{figure}[h]
\begin{centering}
  \includegraphics[height=0.43\linewidth]{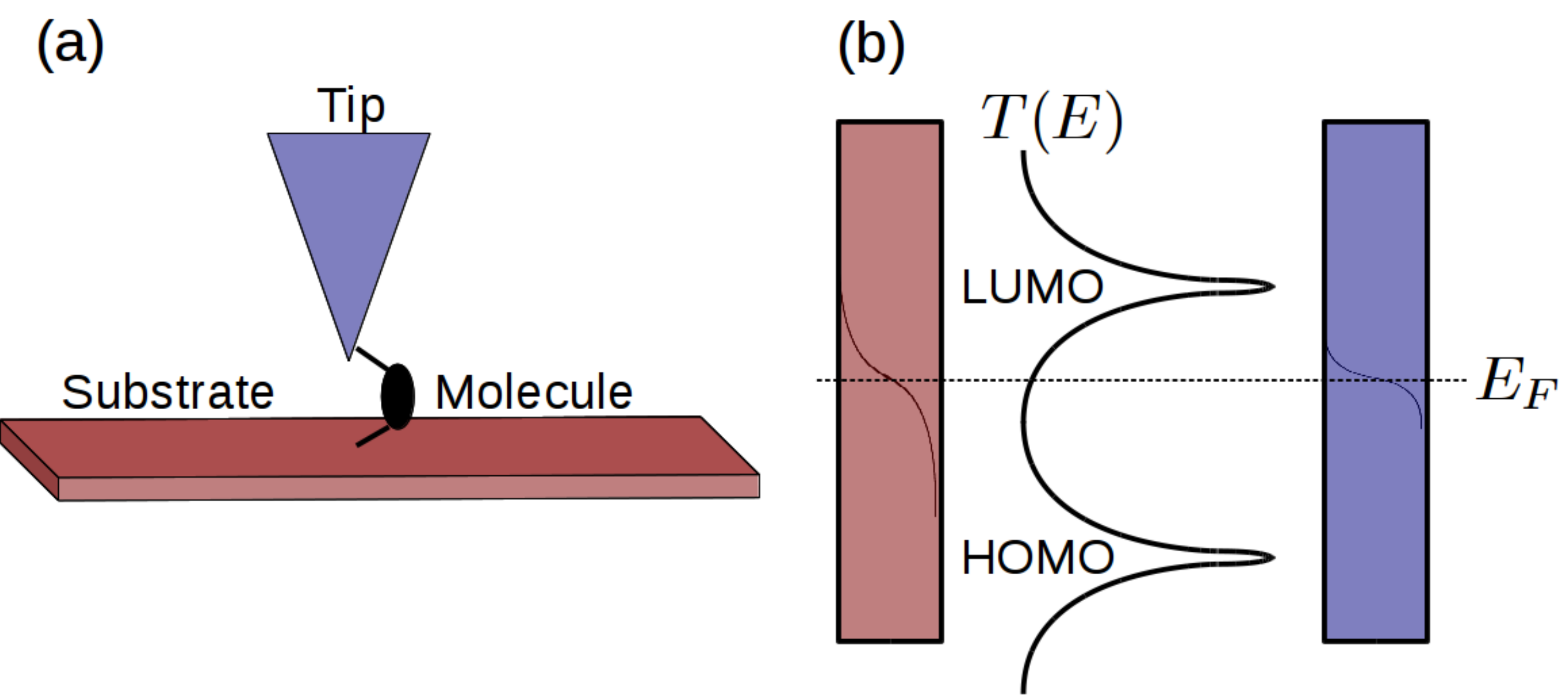}	
	\caption{\label{fig:STM}  (a) Sketch of STM setup used to create single-molecule thermoelectric junctions. 
	(b) Energy diagram of hot (red) and cold (blue) electrode, cf., Fig.~\ref{fig:TEeffect}. Between the electrodes the transmission 
	function of the molecule, $T(E)$, is sketched, which shows peaks at the positions of the HOMO and LUMO.}
\end{centering}
\end{figure}

The Seebeck effect was recently measured in single-molecule junctions, using an STM as sketched in Fig.~\ref{fig:STM}(a), 
see Refs.~\cite{Reddy07,Baheti08,Malen09}. 
A gold substrate is covered with the molecules to be measured, which 
is then heated by passing a current through it (Joule heating), while the STM tip is kept at the ambient temperature.
Conductance measurements are used to verify that a molecule is contacted.
The tip--substrate voltage bias and current amplifier are then replaced by a voltage amplifier, which measures the induced 
thermoelectric voltage as the tip is slowly retracted from the substrate.
When coupled to the tip and substrate contacts, the HOMO and LUMO of the molecules studied in Refs.~\cite{Reddy07,Baheti08,Malen09} 
are far away from the electrode Fermi levels, transport is dominated by elastic (energy-conserving) tunneling and can be described 
by the Landauer-B\"uttiker formalism~\cite{Landauer57,Buettiker86,Appleyard00}. The conductance of the molecular junction is
then proportional to the transmission function at the Fermi energy, $G \propto T(E_F)$.
Often, $T(E)$ is roughly proportional to the molecular density of states and has peaks at the positions of the 
HOMO and LUMO, see Fig.~\ref{fig:STM}(b). As discussed above, the thermoelectric effect relies on electrons above 
and below the Fermi level having different conductances
and in the zero-temperature limit it can be shown that
\begin{align}\label{eq:seebeck_landauer}
	S 	\propto	\frac{1}{T(E_F)} \frac{d T(E_F)}{d E}.
\end{align}
Therefore, the sign of the Seebeck coefficient gives an indication of whether the HOMO or LUMO lies closest to the Fermi level. 

The Seebeck coefficients found in Refs.~\cite{Reddy07,Baheti08,Malen09} were rather modest ($|S| < 10$~$\mu$V/K). 
Much higher values can be expected if either the HOMO or the LUMO lies closer to the Fermi level. 
Furthermore, a more narrow width of the transmission peaks increases the derivative of the transmission function in 
Eq.~(\ref{eq:seebeck_landauer}) and thereby the Seebeck coefficient, provided that the HOMO or LUMO position is adjusted 
appropriately. The width of the transmission peaks is set by the strength of the tunnel couplings to the electrodes and 
therefore a weak coupling is desirable to approach the $\delta$-like conductivity of Mahan and Soho~\cite{Mahan96}. 
That a single-molecule junction can indeed, in theory, operate at ideal efficiency was the conclusion 
of Ref.~\cite{Murphy08}, and Ref.~\cite{Esposito09} showed that such a device has a high efficiency even away from 
equilibrium.
If it proves too difficult to move the HOMO or LUMO close enough to the Fermi level, or if small enough tunnel couplings cannot 
be achieved, an alternative is to use molecules with sharp features in the transmission function,
for example Fano resonances~\cite{Finch09},
\hl{charge-Kondo resonances~\cite{Andergassen11b}}, "transmission supernodes"~\cite{Bergfield10}, or interference-related 
features~\cite{Solomon08,Karlstrom11}, which also result in a strong energy-dependence of the conductance.  

Naturally, a single molecule does not provide enough output power for an actual application as a power converter. 
Instead, a self-assembled molecular monolayer could be used. However, due to e.g., re-arrangement of molecular or 
surface charges or interactions between static molecular dipole moments~\cite{Naaman10,Heimel10,Egger12_mol}, 
or inter-molecular tunneling~\cite{Yaliraki98,Magoga99,Reuter11b,Reuter11}, the transport properties of a molecular monolayer may be rather
different compared with single-molecule devices. In addition, a recent theoretical study~\cite{Leijnse12} showed that Coulomb interactions 
between charge carriers transported through neighboring molecules within a monolayer may significantly broaden the transport
resonances. This might have a negative effect on the thermoelectric efficiency.
 
\subsection{Electron--phonon coupling and phononic heat currents}
Having seen that the electronic properties of molecules hold promise for efficient thermoelectric power conversion, we now 
turn to important contributions to the losses in such devices, namely electron--phonon coupling and coupling between
quantized molecular vibrations and substrate phonon modes in the electrodes. 

Electron--phonon coupling is the coupling between the charge on the molecule and its vibrational motion. Due to 
the electron--phonon coupling, electrons tunneling through the molecule can excite it 
vibrationally~\cite{Park00,Braig03a,Pasupathy04,Osorio07a}. 
In a simple picture, such inelastic processes destroy the $\delta$-like character of the transmission function, 
since electrons can tunnel through the molecule at energies other than the conducting HOMO or LUMO, by either giving off 
excess energy into the molecular vibrations, or by absorbing vibrational energy from them. In a thermoelectric device, the relevant
modes are those with a vibrational energy $\hbar \omega \sim k_B T$. Modes with much larger energies cannot be excited and those with 
much smaller energies do not contribute as much to heat losses. 
Thus, as will be substantiated below based on a simple model, \HL{in a
good thermoelectric molecule} all vibrational modes with $\hbar \omega \sim k_B T$ should have a small 
electron--phonon coupling~\cite{Leijnse10}.

In addition, the molecular vibrational modes couple to substrate phonon modes in the electrodes~\cite{Braig03a}. Essentially, the 
chemical bond between the molecular anchoring groups and the electrodes \HL{acts} as a spring, which can transfer vibrational energy 
from the phonon modes in the hot electrode, into the molecular vibrations, and finally out again into the cold electrode phonons.
The resulting phononic heat current has been analyzed in several theoretical works~\cite{Segal02,Segal06,Galperin07b}.
A strategy to minimize such losses is to choose molecules which form strong chemical bonds to the electrodes (meaning a 
stiff spring), such that all molecular vibrations with a significant amplitude at this bond have frequencies above the highest 
acoustical phonons in the electrodes. A strong chemical bond could be combined with a weak tunnel coupling, as needed to obtain 
sharp electronic resonances, e.g., by connecting the bonding atoms to the rest of the molecule through saturated carbon
atoms, as in a methylene spacer~\cite{Poulsen09}. 
However, a molecule in a transport junction has additional low-energy center-of-mass 
vibrational modes~\cite{Seldenthuis08}, which are likely below the Debye frequency of the electrodes.

\subsubsection{A simple model of a molecule coupled to phonons}
To better understand the concepts discussed above, we now focus on a simple molecular model~\cite{Leijnse10,Galperin08},
which still captures the relevant aspects of electronic and vibrational degrees of freedom.
We model the molecule as a single spin-degenerate orbital level with energy $\epsilon$ and onsite Coulomb interaction $U$ 
and include a single harmonic molecular vibrational mode with frequency $\omega$. 
Furthermore, there is a linear electron--vibration coupling $\lambda_{\rm ph}$ between the electron occupation of the 
orbital and the vibrational coordinate. 
The transport characteristics of this so-called Anderson-Holstein model, when the molecular orbital is tunnel coupled 
to voltage biased source and drain electrodes, has been analyzed in many works, see 
e.g.,~\cite{Flensberg03,Mitra04b,Koch04b,Koch06,Haupt06,Leijnse08a}.
Here we consider instead a thermoelectric junction operated as a heat to electric power converter,
where a hot electrode with temperature $T_h = T + \Delta T$ is grounded (chemical potential $\mu_h = 0$) and a (negative) 
voltage $-V$ is applied to the cold electrode ($\mu_c = V > 0$) with temperature $T_c = T$. 
Note that in an actual device which also makes use of the converted power, the voltage is not applied, 
but rather controlled by the temperature bias and the resistance of the external circuit,
see Fig.~\ref{fig:TEeffect}(d). 
In addition to the electronic tunnel coupling, we include a coupling between the coordinate of the localized molecular 
vibration and a continuum of vibrational modes in the two electrodes. 
Electron tunneling between the molecule and electrode $r = h,c$ is associated with a rate $\Gamma_r$ and the 
corresponding rate at which vibrational excitations "leaks" out from the molecule and into the electrode phonon 
modes is denoted by $\gamma_r$.
As discussed above, the maximum efficiency of energy conversion is expected in the limit of weak 
electron tunneling and, of course, weak coupling between molecular vibrations and electrode phonons, meaning that 
$T \gg \Gamma_r, \gamma_r$.

In the weak coupling regime, the type of master equation introduced in Sec.~\ref{sec:method} can be 
used to calculate the molecular density matrix describing both the electronic and vibrational state.
The interplay between electron and phonon transport is nontrivial, as these 
processes interact via the vibrational distribution on the molecule (which is not necessarily thermalized). 
The electric and the electronic and phononic heat currents can be found as described in Sec.~\ref{sec:method}, and
the efficiency is evaluated from Eq.~(\ref{eq:converter_eta}), $\eta = P / Q_h$, where $P = I V$ with $I = -I_h = I_c$. 
Note that the relevant heat current is $Q_h$ since the loss is given by the heat which must be supplied to keep this electrode hot. 
The distinction is important since there is no conservation of the stationary heat current.
Instead, the first law of thermodynamics guarantees that $P = Q_h + Q_c$, so in fact, to obtain $\eta$, one could completely 
avoid calculating the electric current and instead calculate only $Q_h$ and $Q_c$.

\subsubsection{Optimal bias voltage and level position\label{sec:transport}}
We start by studying the efficiency and output power at fixed thermal bias, here chosen 
to be $\Delta T = T$, as function of $V$ and $\epsilon$.
The efficiency of a single level quantum dot (spinless electrons and no vibrational mode)
was studied in Ref.~\cite{Esposito09}, showing that 
Carnot efficiency is reached in the equilibrium limit of vanishing current.
For vanishing couplings to the phonon mode, $\gamma_r \rightarrow 0$ and $\lambda_{\rm ph} \rightarrow 0$, 
this result remains valid in the non-interacting limit, $U = 0$, as well as for very strong interactions, $U \gg T, \Delta T$,
while the efficiency is slightly reduced in the intermediate regime.

\begin{figure}[t!]
     \includegraphics[height=0.7\linewidth]{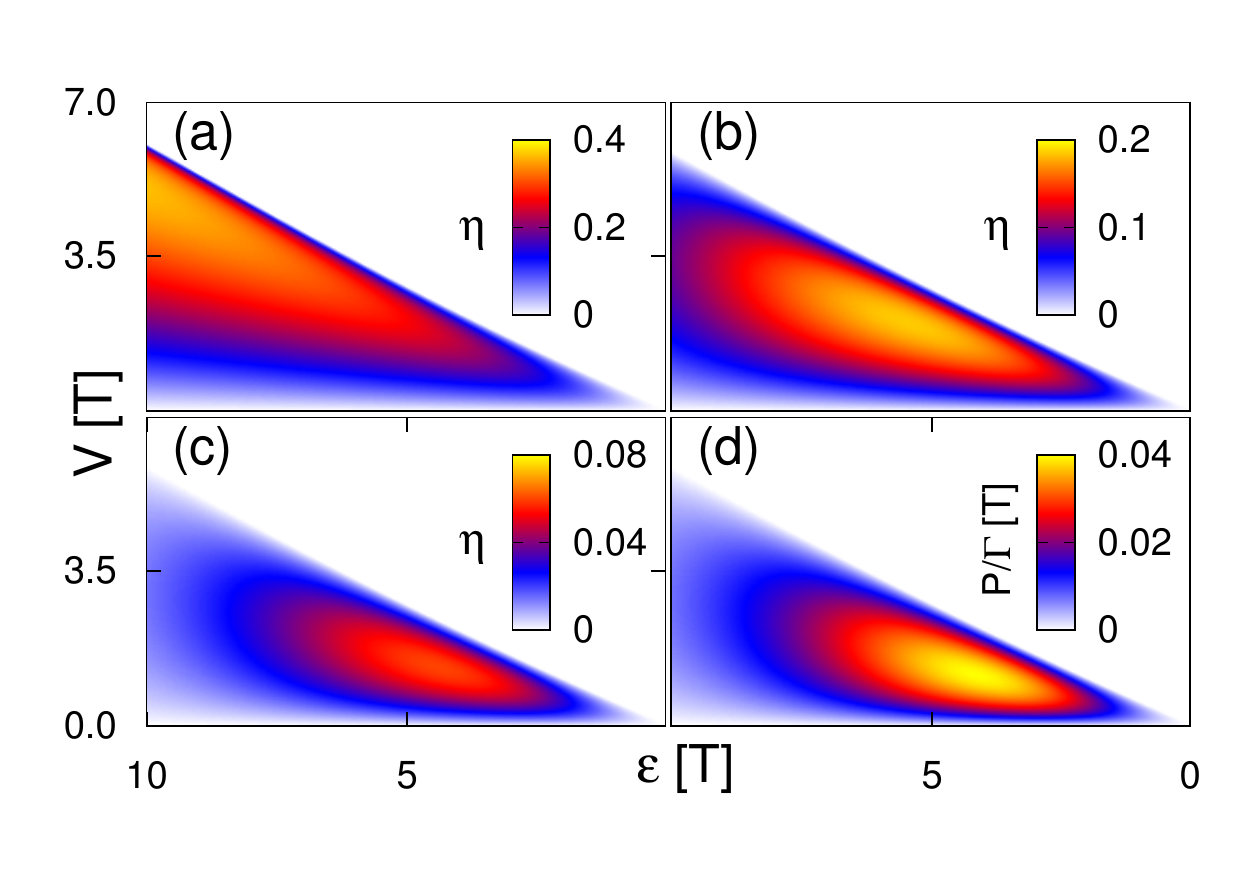}
    \caption{\label{fig:results1} 
	(a)--(c) $\eta$ at $\Delta T = T$, as function of $V$ and $\epsilon$ for increasing coupling to substrate phonons, 
	$\gamma = 0$ (a), $\gamma = \Gamma / 10$ (b), $\gamma = \Gamma$ (c). In all plots
	$\lambda_{\rm ph} = 1$, $\omega = T$, $U = 10 T$ and the couplings are symmetric, 
	$\gamma_h = \gamma_c = \gamma$, $\Gamma_h = \Gamma_c = \Gamma$.
	(d) $P$ as function of $V$ and $\epsilon$ for the parameters used in (b) (the power depends only weakly on $\gamma$).}
\end{figure}

With finite $\lambda_{\rm ph}$, but keeping $\gamma_r = 0$, the efficiency
is decreased and never reaches the Carnot value, see Fig.~\ref{fig:results1}(a).
In fact, $\eta$ vanishes close to the zero electric current line (boundary of the white area), 
the reason being that, in contrast to the single-level case, the heat current does not vanish completely when the charge current does.
Inside the white area the current has been reversed by a too large voltage bias and flows from 
high- to low-biased electrode and therefore does not accomplish any useful electric 
work (note that this regime cannot be reached in the thermoelectric circuit of Fig.~\ref{fig:TEeffect}(d)).
The maximal efficiency is reached when the level is far above the Fermi edges of both leads, 
where electron transport involves very few thermally excited states 
in the heated electrode and electron-induced vibrational excitations are exponentially suppressed, minimizing 
electronic heat loss.
However, in this regime the current is highly suppressed, leading to a small 
output power, see Fig.~\ref{fig:results1}(d).
In addition, even a small coupling to the substrate phonons, $\gamma = \Gamma / 10$ in Fig.~\ref{fig:results1}(b), 
drastically decreases the efficiency in this low-current regime, while having a smaller effect in the regime 
where the current is larger ($\epsilon$ is smaller). 
Thus, even a weak coupling to substrate phonon modes, $\gamma \ll \Gamma$, drastically changes the ideal 
operating conditions for maximum $\eta$ by introducing a heat loss which depends only weakly on $\epsilon$ and $V$.
When the coupling to the substrate phonons becomes comparable to the tunnel coupling, $\gamma \approx \Gamma$ 
in Fig.~\ref{fig:results1}(c), the efficiency is significantly decreased also in the high current regime.

\subsubsection{Temperature dependence and molecular heating}
\begin{figure}[t!]
  \includegraphics[height=0.4\linewidth]{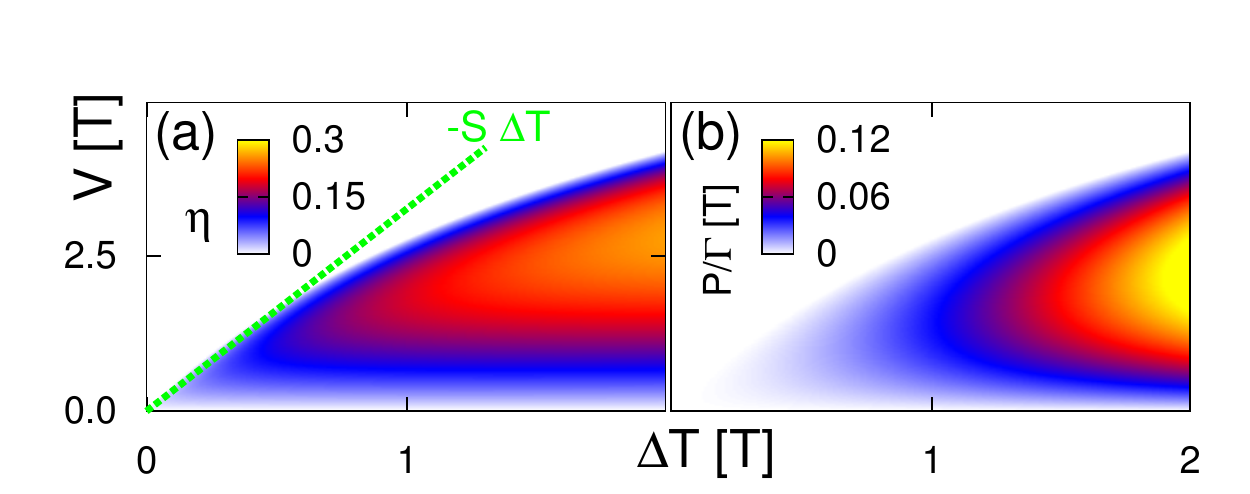}
  \caption{\label{fig:results2} 
    (a) $\eta$ and (b) $P$ 
    as a function of $V$ and $\Delta T$, with $\epsilon = 5 T$ and other parameters as 
    in Fig.~\ref{fig:results1}(b).
    }
\end{figure}

Next we fix the level position to a value with both large power and efficiency,
$\epsilon = 5 T$, and vary instead $V$ and $\Delta T$. 
The resulting efficiency and output power is shown in Figs.~\ref{fig:results2}(a) and (b) for the same parameters as in 
Fig.~\ref{fig:results1}(b). 
As above, a too large voltage bias compared to the temperature bias reverses the current and 
no useful electric work is accomplished (white areas).
The non-linear thermopower can be defined through $V = -S(\Delta T) \Delta T$ at $I = 0$, i.e., 
$V$ is the finite voltage needed to compensate the temperature bias and give zero electric current. 
As expected, both efficiency and power are increased by an increased temperature bias. 

Comparing Fig.~\ref{fig:results2}(a) and (b), we see that for a given temperature bias, 
maximum efficiency and maximum output power is achieved at almost the same voltage bias.
The reason is seen from the relation $P=\eta~( Q_h^{(e)} + Q_h^{(p)}) $,
where $Q_h^{(e)}$ ($Q_h^{(p)}$) is the electron (phonon) contribution to the heat current.
Since $Q_h^{(p)}$ only has a weak (indirect) dependence on the voltage bias,
$\eta$ and $P$ can be simultaneously maximized by adjusting the bias when the phonon heat loss 
dominates ($\gamma \gg \Gamma$). As Fig.~\ref{fig:results2} shows, this holds approximately 
also when $\gamma \sim \Gamma$.

Ref.~\cite{Leijnse10} investigated also the effects of $\lambda_{\rm ph} > 1$, which further reduces both $\eta$ and $P$, 
and $\lambda_{\rm ph} < 1$, which increases $\eta$. Furthermore, both low- and high-energy 
vibrational modes ($\omega \ll T$ or $\omega \gg T$), have a much smaller effect on $\eta$. 
A low frequency mode can essentially be seen as a broadening of the electronic resonance 
of width $\sim \omega \lambda_{\rm ph}^2$. 
Almost all decrease in efficiency in this case comes from the coupling to substrate phonon modes. 
A high-frequency mode, on the other hand, cannot be excited and therefore does not contribute at all to electron or heat transport.

%% file: include-intro-pumping.tex

\section{Time-dependent driving and adiabatic pumping} 
\label{sec:pumping} 

Charge or heat currents in an electric conductor are typically generated
by imposing a voltage or temperature gradient between the electrodes of
a device, as in the case of Sec.\ref{sec:heat}.
However, in a mesoscopic conductor, a dc-current can flow even without
external gradients if some parameters of the system are periodically
modulated in time~\cite{Thouless83}. 
When the modulation is slow compared to the characteristic time scales
of the system, 
the transport mechanism is called {\em adiabatic
pumping}. 
In this case, the pumped charge depends only on the geometry of the
pumping cycle in parameter space, i.e. it is of geometric
nature~\cite{Avron00,Makhlin01b}.
The interest in adiabatic pumping \hl{has various motivations.
For one, in} appropriate modulation setups~\cite{Niu84,Avron01}, the charge
pumped per period is quantized in units of the electron charge.
Adiabatic charge pumps are therefore promising candidates for a \hl{very}
precise current standard for metrology~\cite{Pekola_rev}. Furthermore,
the controlled emission of single charges~\cite{Feve07} is of interest
for quantum information and electronic analogs of quantum \hl{optical}
effects~\cite{Bocquillon13}.
In the opposite case where the transferred charge is not necessarily
quantized, the pumping mechanism can be dominated by quantum
effects~\cite{Zhou99}. This yields the possibility \HL{of revealing} internal
properties of the device which are not visible from standard transport
spectroscopy through stationary setups. We will discuss examples for
this \textit{adiabatic transport spectroscopy} in the following
sections.

In the last years, adiabatic pumping in solid-state devices has been widely studied both
experimentally~\cite{Pothier92,Switkes99,Watson03,Fletcher03,Chorley12}
and theoretically. As long as interactions can be treated on a
self-consistent mean-field level~\cite{Buettiker93a},
adiabatic pumping is well described by Brouwer's
theory~\cite{Brouwer98}, which is based on a generalization of the
scattering matrix approach for time-dependent
phenomena~\cite{Buttiker94}. This formalism has been applied to study
several aspects of pumping in non-interacting systems, such as
dissipation and noise~\cite{Makhlin01b,Moskalets02}, and the possibility
of spin-pumping~\cite{Mucciolo02,Governale03}. Further works dealt with
different setups, including normal metal-superconductor hybrid
structures~\cite{Wang01,Blaauboer02,Taddei04}, pumping by surface
acoustic waves~\cite{Levinson00,Entin01}, and graphene-based quantum
pumps~\cite{Prada09}.
Heat currents have been considered as well in non-interacting electronic
quantum-pumps, both in the limits of adiabatic~\cite{Moskalets02} and
non-adiabatic driving~\cite{Arrachea07,Rey07,Moskalets04}. 

For systems dominated by a strong Coulomb interaction, the mean-field
approach breaks down and new formulations are necessary to describe
pumping. Several studies addressed interaction effects in specific
setups and regimes. Pumping in interacting quantum wires has been
discussed in Refs.~\cite{Citro03,Das05}. Pumping through open quantum
dots was addressed in Ref.~\cite{Aleiner98,Brouwer05} by employing
bosonization techniques, while the Keldysh Green's function approach has
been applied to investigate pumping in interacting quantum
dots~\cite{Splettstoesser05,Sela06}, including  the Kondo
regime~\cite{Aono04,Schiller08,Arrachea08}. 
A diagrammatic real-time approach~\cite{Splettstoesser06} was used to
investigate several aspects of adiabatic pumping through interacting
quantum-dot systems~\cite{Cavaliere09,Splettstoesser08,Winkler09,Reckermann10a,Riwar10,Riwar12}
weakly coupled to the leads, and served as the basis for non-equilibrium
renormalization group studies that treat the tunneling
non-perturbatively~\cite{Kashuba12}.

Interestingly, adiabatic pumping effects, i.e. the occurrence of a finite flux in a preferred direction in response to a slow periodic or random zero-mean perturbation, are relevant not only in solid-state mesoscopic devices, but also for ion-channels in cell membranes~\cite{Astumian03}, enzymatic reactions~\cite{Sinitsyn07} and, stochastic kinetics in general~\cite{Sinitsyn09}.
\subsection{Geometric properties of adiabatic pumping}
\label{sec:methods:adiabatic}
In terms of the master equation approach outlined in Sec.~\ref{sec:method}, adiabatic pumping is described by the contribution  to the current to first order in the modulation frequency,  $R^{(1)}_{t}$ ($R\in\{I_{r},Q_{r},J_{r}\}$). 
To emphasize its geometric aspects it is useful to introduce auxiliary vector fields
in parameter space~\cite{Avron00}. The key observation is that
$R^{(1)}_{t}$ is directly related to the time derivative of the
instantaneous occupation probabilities,
\begin{equation}
\label{eq:A1}
R^{(1)}_{t}= {\rm Tr}\left\{ \boldsymbol{\mathcal{W}}^{R}_t
\tilde{\W}_{t}^{-1}\tfrac{d}{dt}\P_{t}^{(0)}\right\}\equiv
\boldsymbol{\varphi}^{R}\tfrac{d}{dt}\P_{t}^{(0)}, 
\end{equation}
where $\tilde{\W}_{t}^{-1}$ is a pseudo-inverse of the evolution
Kernel~\cite{Calvo12a}. In the second identity we introduced the
\hl{vector-valued} response coefficients $\boldsymbol{\varphi}^{R}$,
\hl{which describes for $R = I_{r}$,$Q_{r}$,$J_{r}$,}
the rate at which charge, heat and spin respectively, \hl{is} transferred to lead $r$ due to a change in the occupation probabilities. 
The average current pumped per period $\T={2\pi}/{\Omega}$, can
then be written as a line-integral over a closed contour $\partial
\Sigma$ in the space of the driving parameters~\cite{Calvo12a,Ren10}
\begin{equation}
\label{eq:chargeA}
\bar{R}^{(1)}\equiv \T^{-1} \int_{0}^{\T}dt\,
R^{(1)}_{t}=\T^{-1}\oint_{\partial \Sigma} d\bm{u} \cdot
\bm{\mathcal{A}}_R(\bm{u}), 
\end{equation}
with
$\bm{\mathcal{A}}_R =
\sum_{\xi}\varphi_{\xi}^{R}(\bm{u})
\bm{\nabla}P_{\xi}^{(0)}(\bm{u})$. The bar indicates that the average
over one driving period has been taken. Here $\bm{u}=\sum_i
u_i\bm{e}_{i}$ is the ``position'' vector in parameter space and
$\bm{\nabla}=\sum_{i}\bm{e}_i\partial_{u_{i}}$. The vector field
$\bm{\mathcal{A}}_R$ can be interpreted as a pseudovector potential
defined in the space of the driving parameters~\cite{Calvo12a}, and its
components are directly related to the concept of
emissivity~\cite{Buttiker94,Sela06}.
From Eq.~\eqref{eq:chargeA} it follows directly that at least two
independent driving parameters are required to have a non vanishing
pumped current, $\bar{R}^{(1)}\neq0$. In the case of two-parameters
pumping, $\bm{u}=\{u_{1},u_{2}\}$, using Stoke's theorem,
Eq.~\eqref{eq:chargeA} can be written as the surface integral
\begin{align}
\label{eq:Qxi:B}
\bar{R}^{(1)}=\T^{-1}
 \iint_{\Sigma}d\boldsymbol{S}\cdot\boldsymbol{\mathcal{B}}_{R}(\boldsymbol{u}),
\end{align}
where
$\boldsymbol{\mathcal{B}}_{R}(\boldsymbol{u})=\boldsymbol{\nabla}\times
\boldsymbol{\mathcal{A}}_{R}(\boldsymbol{u})$ and $d\boldsymbol{S}$ is
the directed surface element in parameter space. The average current
$\bar{R}^{(1)}$ can then be seen as the flux of the pseudo-magnetic
field $\boldsymbol{\mathcal{B}}_{R}$. The advantage of this
representation is that $\boldsymbol{\mathcal{B}}_{\R}$ anticipates the
conditions for finite pumping without referring to the specific details
of the modulation~\cite{Calvo12a}. 

The geometric interpretation holds however only to first order in the
driving frequency and cannot be generalized to higher-order
contributions, which depend sensitively on how the pumping orbit is
traversed.

%% file: include-pumping-js.tex

\subsection{Adiabatic transport spectroscopy}
\label{sec:pumpch}

\begin{figure}
\begin{centering}
\includegraphics[width=0.45\textwidth]{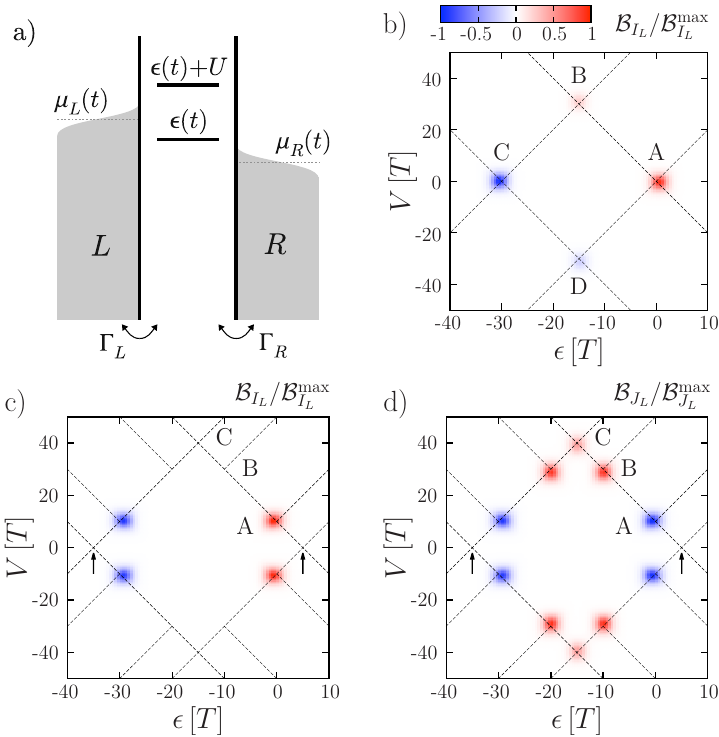}
\caption{ (a) Scheme of the transport device. (b)
 Stability diagram for the pumped charge: gate and bias map of the
 normalized pseudo-magnetic field for $\Delta = 0$ and $\lambda =
 0.25$. Bottom panels: Normalized pseudo-magnetic fields for $\Delta =
 10T$ associated to the pumped charge (c) and spin (d) for $\lambda =
 0$. The chosen Coulomb interaction is $U=30T$.}
\label{fig:pump1}
\end{centering}
\end{figure}

In this section we will show how the  geometric interpretation of
adiabatic pumping can be used for the analysis of adiabatic transport
spectroscopy. As an example we study a single-level quantum dot in a
non-equilibrium regime induced by the modulation of the dot's level
position and the applied bias around a \emph{finite value}. This situation
has been addressed both in non-interacting
systems~\cite{Moskalets04,Entin02} within the context of the scattering
matrix theory and, more recently, in a strongly interacting quantum
dot weakly coupled to leads~\cite{Reckermann10a,Calvo12a}. We will focus on this last example and show that pumping is interaction-induced and can be used as a spectroscopic tool
to access information on spin degeneracy and junction asymmetry in the
quantum dot, \hl{complementing standard \hl{dc-spectroscopy} (``stability diagrams'').}
Based on Sec.~\ref{sec:methods:adiabatic}, we describe the interplay
between interaction and non-equilibrium effects in terms of the
pseudo-magnetic fields associated to the charge and spin currents. To
\hl{this} end, we consider the transport setup shown in Fig.~\ref{fig:pump1}(a). The
quantum dot is described by the Hamiltonian $H_D(t) = \sum_\sigma
(\epsilon(t) - \sigma \Delta/2) n_\sigma+U n_\uparrow n_\downarrow$,
where $\Delta$ is the Zeeman splitting due to an applied external
magnetic field and $U$ is the onsite Coulomb energy. The level energy of
the dot $\epsilon(t)$ is driven by a time-dependent gate voltage while
the electro-chemical potentials in the reservoirs are controlled by the
bias, i.e. $\mu_r(t)=\pm V(t)/2$ for $r=\{L,R\}$. These two voltages are
slowly driven \hl{at frequency $\Omega$} in a circular orbit around a working point
$(\bar{\epsilon},\bar{V})$.

Following Sec.~\ref{sec:method}, the charge and spin currents $R \in
\{I_r, J_r\}$ are calculated in lowest order in the coupling to the
leads, and we will focus on the adiabatic correction (linear order in
the modulation frequency) to the instantaneous solution. After a full
cycle of the modulation, the net amount of charge/spin transferred to
the $r$-lead is given by the instantaneous current flowing in response
to the finite, time-dependent bias and the additional pumped
charge/spin, $\mathcal{N}_R = \mathcal{T}\bar{R}^{(1)}$, generated by
the delayed response. Experimentally, this adiabatic contribution can be
extracted from the total current by using a lock-in technique.
According to Sec.~\ref{sec:methods:adiabatic}, it is possible to
represent the average current, and hence $\mathcal{N}_R$, in terms of a
pseudo-magnetic field which in this case \hl{reads as}
\begin{equation}
\boldsymbol{\mathcal{B}}_R({\bm u}) =
\boldsymbol{\nabla}\varphi_n^R \times
\boldsymbol{\nabla}\braket{n} + \boldsymbol{\nabla}\varphi_s^R \times
\boldsymbol{\nabla} \braket{S_z},
\label{eq:pseudob}
\end{equation}
with the driving parameters ${\bm u} = (\epsilon, V)$. Instead of
occupation probabilities, Eq.~(\ref{eq:A1}), we here use the
instantaneous dot's average charge $\braket{n}_t$ and spin
$\braket{S_z}_t$, respectively. The response coefficients
$\varphi_{n,t}^R$ and $\varphi_{s,t}^R$ determine the amount of
charge/spin leaving the $r$-lead when either the average charge or spin
in the dot is changed in response to the driving.

In the non-interacting limit ($U=0$), the response coefficients turn out to be
independent of the driving parameters and hence they yield a zero
pseudo-magnetic field such that the pumped charge and spin completely
vanish. Adiabatic pumping in the discussed regime is hence
\textit{interaction-induced}.

The finite pseudo-magnetic field generated by a large Coulomb charging
energy ($U \gg T$) and zero magnetic field is shown in
Fig.~\ref{fig:pump1}(b). A finite pumped charge occurs in peaks located
around the meeting point of two dot level resonance lines (dashed lines
in the figure); for the remaining regions $\mathcal{B}_{I_r}$ (and
therefore the pumped charge) vanishes for the following reasons: (i)
When the driving is far away from any resonance line, the occupation in
the dot remains constant, and its time-derivative is hence zero. (ii)
When the full driving trajectory only crosses a single resonance line,
the response coefficient and the occupation number depend on the same
effective parameter and hence their gradients are parallel.

The low-bias peaks (labeled by A and C), displayed in the ``stability
diagram for the pumped charge'' in Fig.~\ref{fig:pump1}(b) are
dominant, while the high-bias peaks (B and D) \hl{only} emerge when \hl{a non-zero} asymmetry
$\lambda = (\Gamma_L-\Gamma_R)/\Gamma$ between left and right contacts
is \hl{present}. Moreover, the adiabatic pumping can here be used to
\hl{directly} identify the \hl{nature of the coupling asymmetry:}
the peaks A and B in Figs.~\ref{fig:pump1}(b), \hl{are}
related to each other by
\begin{equation}
 \mathcal{B}_{I_L}^{({\rm B})}(\epsilon,V) = \lambda \,
  \mathcal{B}_{I_L}^{({\rm A})}(V/2,2\epsilon).
\label{eq:highlow}
\end{equation}
Since $|\lambda|<1$, the magnitude of the peak B is always smaller
than the one of peak A and its \hl{\emph{sign}} is determined by the sign of
\hl{the coupling asymmetry} $\lambda$. For any two modulation curves centered around these points, a
change of variables allows to write $\mathcal{N}_{I_L}^{({\rm B})} =
\lambda \, \mathcal{N}_{I_L}^{({\rm A})}$, such that the {\em mere}
presence of a pumped charge in the high-bias regime indicates an
asymmetric coupling to the leads. Interestingly, this simple relation
\hl{can be used for} a direct quantitative estimation of the coupling
asymmetry by \hl{dividing the independently measured values of pumped charges at the points B and A.}

Note that in contrast to metrological applications of adiabatic pumping,
the transferred charge is never quantized here. When the area enclosed
by the modulation in parameter space includes a full peak, the pumped
charge reaches however a plateau whose maximum value, corresponding to
$\lambda=0$, is $1/6$ in units of the electronic charge.

In the presence of a finite \hl{external} magnetic field, Fig.~\ref{fig:pump1}(c),
the resonances of Fig.~\ref{fig:pump1}(b) are split into further
well-separated peaks. Interestingly, regardless of the value of
$\lambda$, there is no peak \hl{at the zero-bias crossing points} (black
arrows). In this regime of the driving, the vector potentials for the
charge and spin currents are irrotational, such that integration over a
closed trajectory yields zero pumping. The reason for this is that a
finite pumped charge requires not only a modulation encircling the
meeting point of two resonance lines but also a change in the spin
degeneracy of the ground state~\cite{Reckermann10a}. This requirement
becomes evident for the crossing at $V = \Delta$, where the spin
degeneracy is effectively recovered through the applied bias. \hl{Charge pumping} is
hence also an indicator of the spin degeneracy occuring in the level
spectrum of a quantum dot.

\subsection{Spin pumping}
\label{sec:Bfinite}
In addition to charge pumping, the possibility to pump spin is an
interesting option due to its significant robustness against environment
decoherence. As fundamental elements in the realization of spin-based
electronics, spin pumps could operate in a wide range of setups
including `turnstile spin pumps', i.e., quantum dots in presence of an
external magnetic field where the spin current flows in response to a
periodic modulation of the confining
potentials~\cite{Watson03,Mucciolo02}, or combinations of driving
parameters like the coupling to the leads and the amplitude of the
magnetic field~\cite{Blaauboer05}. Other interesting examples exploit
the spin polarization in the reservoirs, like heterostructures
consisting of normal metal and precessing ferromagnetic leads acting as
a spin battery~\cite{Brataas02,Winkler13}, or quantum systems where the
spin-orbit coupling is at the core of the
phenomenon~\cite{Governale03,Rojek13}. In a double quantum dot coupled
to normal metal and ferromagnetic contacts, pure spin pumping is
obtained when modulating the level positions of the coupled
dots~\cite{Riwar10}.

\hl{Returning} now to the setup of Fig.~\ref{fig:pump1}(a), an external
magnetic field ($\Delta \gg T$) generates a spin current $J_{r,t}^{(1)}$
that flows through the dot in response to the time-dependent
modulation. The above adiabatic spectroscopy of the resonance peaks for
the pumped charge can also be extended to the spin degree of freedom,
where one could test how the pumped spin relates to the pumped charge at
different \hl{biases}. To \hl{this} end, we consider an external magnetic field that
induces a non-zero adiabatic spin current flowing through the dot.

In Fig.~\ref{fig:pump1}(d) we show stability diagrams for the pumped
spin in the presence of a finite magnetic field, $\Delta\neq 0$. The
difference between charge and spin fields is particularly pronounced for
\hl{a symmetric junction ($\lambda = 0$)}, where {\em pure} spin pumping occurs in the high-bias regime:
the pumped charge peaks B and D at high bias vanish exactly while those related to the pumped spin remain finite.

\begin{figure}
 \begin{center}
   \includegraphics[width=0.45\textwidth]{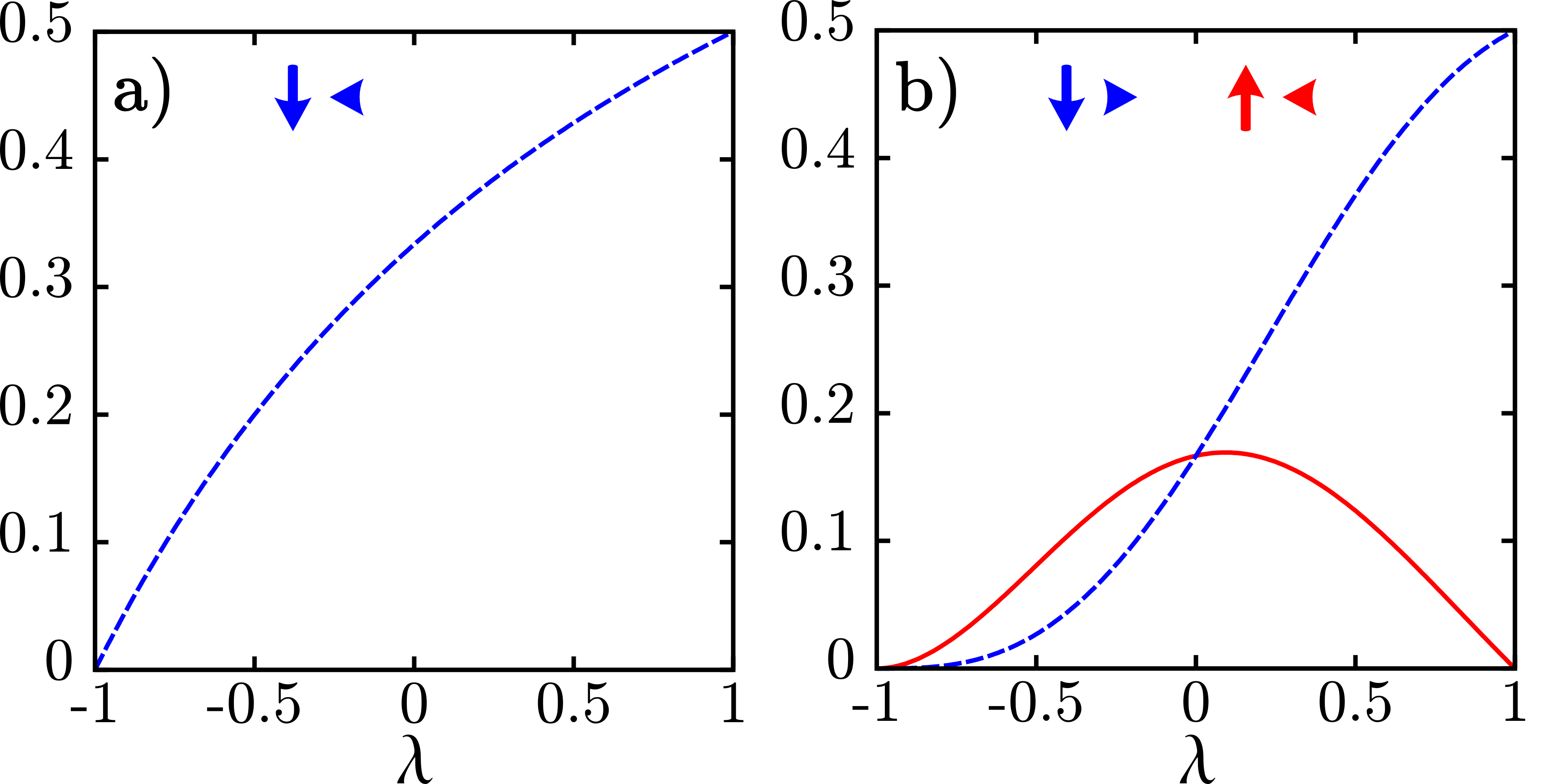}
   \caption{ Absolute value for the spin-resolved pumped
  charge
  $\mathcal{N}_{I_L^\sigma}=(\mathcal{N}_{I_L}+2\sigma\mathcal{N}_{J_L})/2$
  for peaks A and B in Figs.~\ref{fig:pump1}(c) and (d). Directions for
  spin-up (solid red) and spin-down (dashed blue) pumped charge are
  denoted by arrows.}
   \label{fig:pump2}
 \end{center}
\end{figure}

In Fig.~\ref{fig:pump2} we show the spin-resolved pumped charge
associated to the peaks of the pseudo-magnetic fields in
Figs.~\ref{fig:pump1}(c) and (d). Its sign, independent of $\lambda$,
fixes the direction of the overall current after one cycle of the
driving, as indicated by the arrows in the figure.
In panel (a), the pumped charge \hl{at peak A
is fully polarized:
transport of spin down from the right to the left lead causes the net flow of spin.} Panel (b)
shows the pumped charge \hl{at} peak B. Strikingly, the pumped charge
for the two spin orientations always flow in opposite directions. In the
symmetric case $\lambda = 0$, their magnitudes are exactly the same,
such that the total transported charge cancels out while the pumped spin
remains finite. The same qualitative behavior is found also for the
currents around the peak C.

Finally, this analysis can be extended to the remaining peaks in
Figs.~\ref{fig:pump1}(b)-(d) by using the symmetries of the
pseudo-magnetic fields around the particle-hole symmetry point
$(\bar{\epsilon},\bar{V}) = (-U/2,0)$. Specifically, we find an
antisymmetric shape for the pseudo-magnetic field associated to the
pumped charge, i.e. $\mathcal{B}_{I_L}(-{\bm u}) =
-\mathcal{B}_{I_L}({\bm u})$ whereas the one for the pumped spin is
symmetric, i.e. $\mathcal{B}_{J_L}(-{\bm u}) = \mathcal{B}_{J_L}({\bm
u})$.

%% file: include-heat-pumping.tex

\section{Pumping heat with a driven double quantum dot\label{sec:heatpumping}}
In Sec.~\ref{sec:heat} we considered the performance of a molecular quantum dot as thermoelectric engine in the stationary state. The study of quantum-dot devices as nanoscale engines can be extended to the case of time-dependently driven systems, where both the specific properties of the device and the external driving can now be exploited to perform useful work (e.g. by moving an electron from a lower to a higher chemical potential) or to cool a reservoir.  
In particular, the time-dependent modulation of a quantum-dot setup allows to realize mesoscopic analogs of cyclic heat engines that are sequentially coupled and decoupled to hot and cold reservoirs \cite{Esposito10}.

\begin{figure}[tbh]
\centering
\includegraphics[width=0.9\columnwidth]{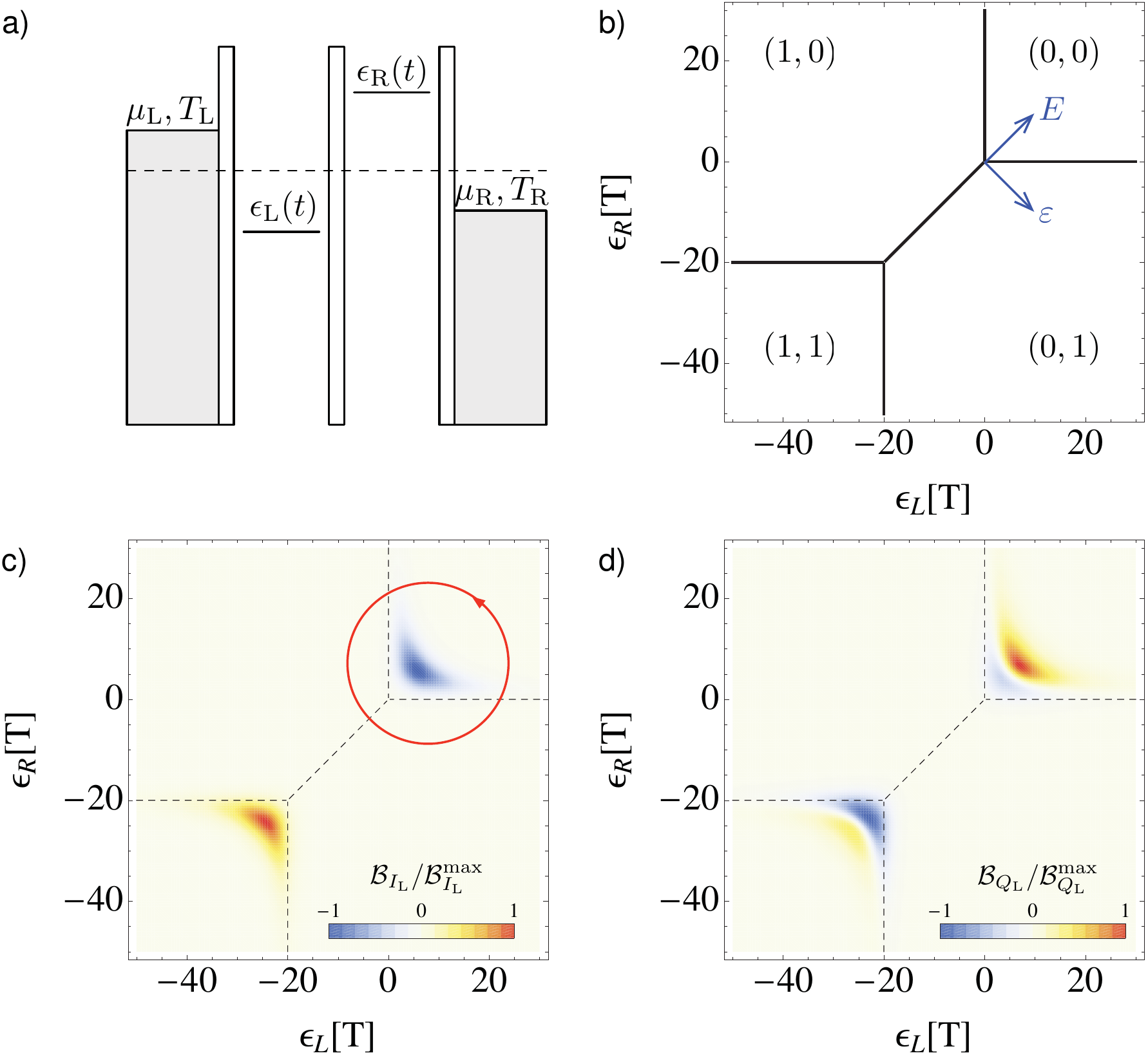}
\caption{
(a) Sketch of the potential landscape of a double-dot setup. (b) Stability diagram of the double dot for the case $V=0$, and $\Delta T=0$: black  lines indicate the borders of the stability regions for negligible inter-dot \HL{coupling.} The system of coordinates formed by the detuning $\e=\epsilon_{\rm L}-\epsilon_{\rm R}$ and the mean energy $E=(\epsilon_{\rm L}+\epsilon_{\rm R})/2$ is also shown.
(c) Color scale plot of the pseudo magnetic field associated with charge pumping $\mathcal{B}_{I_{\rm L}}$ (normalized to its maximum value) and sketch of a possible pumping trajectory in parameter space. (d)  Color scale plot of the pseudo magnetic field associated with heat pumping $\mathcal{B}_{Q_{\rm L}}$. In panels (b)-(d): $U=20 T$, $t_{\rm c}=10 T$, \HL{$\Gamma \ll T $}.}
\label{fig_setup}
\end{figure}

Here, we focus on the case of a double-dot pump (see Fig.~\ref{fig_setup}a) which, differently from the single-dot setup of Sec.~\ref{sec:pumpch}, permits pumping one electron per cycle even in the presence of a finite bias~\cite{Pothier92,Chorley12}.
\hl{Moreover, it allows for the implementation of an efficient effective decoupling scheme from the reservoirs.}
The double dot is described by the Hamiltonian
\hl{
${H}_\mathrm{D} =  \sum_{r} \epsilon_{r}{n}_{r}+U {n}_\mathrm{L}{n}_\mathrm{R}
+\frac{U'}2\!\!\sum_{r}\!\!{n}_{r}({n}_{r}-1) 
 -\frac{t_{\rm c}}{2}\sum_{\sigma} ({d}_{\mathrm{L}\sigma}^\dagger{d}_{\mathrm{R}\sigma}^{}+\mathrm{h.c.} ),
$
}
where ${n}_r
=\sum_{\sigma}{d}_{r\sigma}^\dagger{d}_{r\sigma}^{}$ is the occupation number operator of dot $r=\mathrm{L},\mathrm{R}$ and $\epsilon_{r}$ the corresponding single-particle energy, which can be modulated in time by external gates $\epsilon_r=\epsilon_r(t)$. Interactions between electrons in the same or in different dots are accounted for by $U'$ and $U$, while $t_{\rm c}$ \HL{represents} the inter-dot tunneling amplitude.
In the following, we will assume the onsite interaction $U'$ to be the largest energy scale in the system -- so that each dot can be at most singly occupied -- and consider the case of strong inter-dot coupling $t_{\rm c}\gg\Gamma$. \hl{Finally, we assume that} the system is symmetrically tunnel coupled to two non-interacting electronic reservoirs,  
which are in equilibrium at the temperatures $T_{\rm L}=T+\Delta T$ and $T_{\rm R}=T$ and chemical potentials $\mu_{\rm L}=-\mu_{\rm R}=V/2$.

\hl{The standard ``honeycomb''} stability diagram of the double dot is shown in Fig.~\ref{fig_setup}(b), and identifies regions of different equilibrium occupation numbers for the two dots, as a function of the energies $\epsilon_{\rm L}$ and $\epsilon_{\rm R}$.  The regions where three charge states are degenerate are named {\em triple points}. 
The pump is operated by applying a sinusoidal voltage to the local gates of the two dots with a $\pi/2$ phase shift between them, which forces the state of the system to follow \hl{a circular orbit} in parameter space, see e.g. Fig.~\ref{fig_setup}(c). 

\subsection{Quantized charge and heat pumping}
\begin{figure}
\includegraphics[width=0.9\columnwidth]{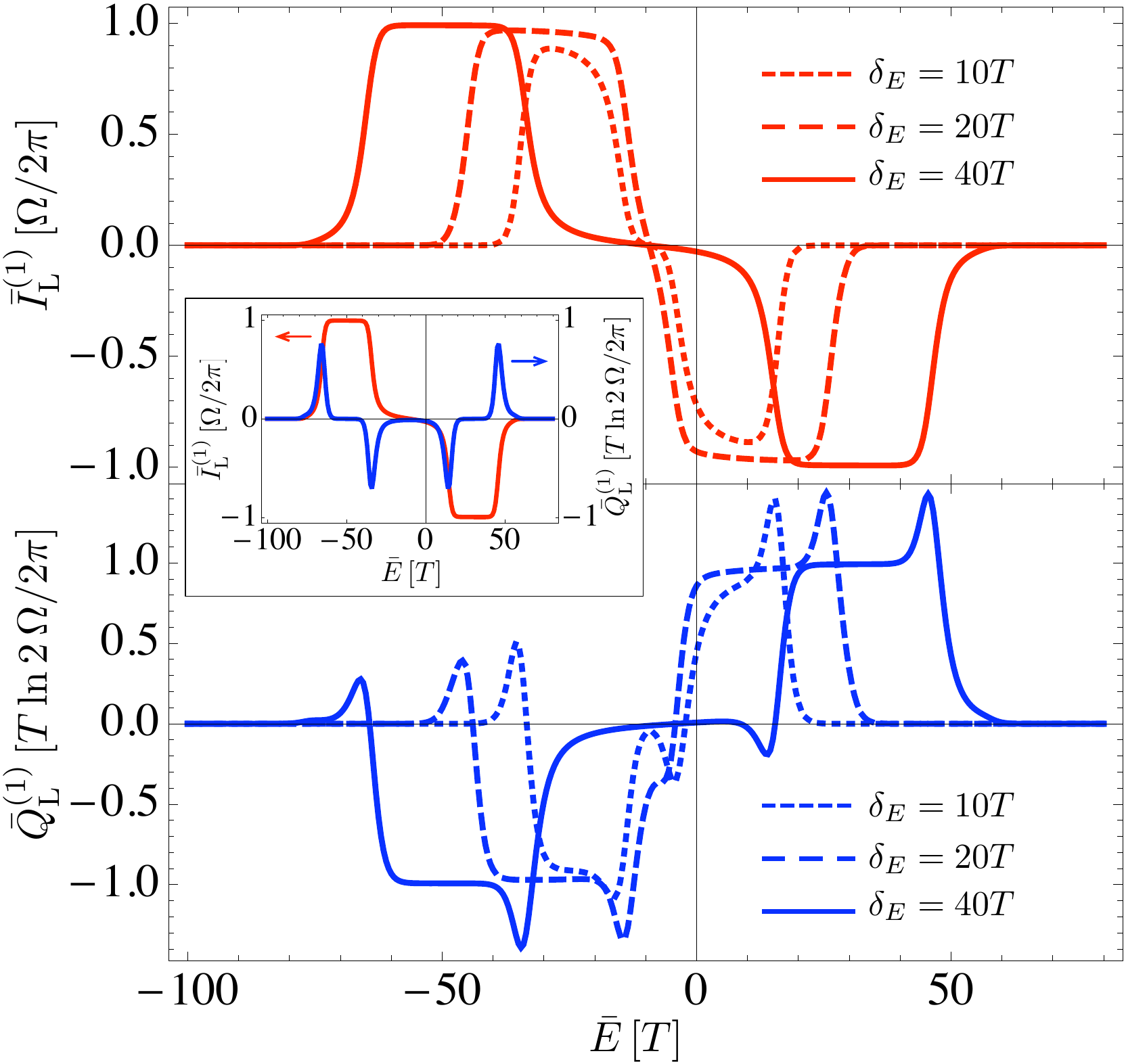}
\caption{ Upper panel:  Average pumped charge-current plotted as a function of  the mean energy $\bar{E}$, for three different pumping orbits.
The pumping cycle is defined by: $E(t)=\bar{E}+\delta_{E}\sin(\Omega t)$ and $\e(t)=2\delta_{E}\cos(\Omega t)$, where $E$ is the mean energy and  \HL{$\delta_{\e}$ the driving amplitude} (see Fig.~\ref{fig_setup}(b)).
Lower panel: same as above, but for the pumped heat-current. Inset: Average pumped charge and heat current for the case of a fully spin-polarized system. In all panels:  $\Delta T=0$, $V=0$ and $U=20 T$, $t_{\rm c}=10 T$,  \HL{$\Gamma \ll T$}.}
\label{fig:currents-omega1-eq}
\end{figure}

We consider first the case of pure {\em adiabatic pumping}, 
($\Delta T=0$, $V=0$ and $\Omega\to0$), where the only relevant contributions to the currents are those to first order in the driving frequency $\Omega$, i.e., $R^{(1)}(t)$.  As discussed in Sec.~\ref{sec:methods:adiabatic}, the average current $\bar{R}^{(1)}$ has a geometric interpretation and it can be expressed as the flux of the pseudo magnetic field $\bm{\mathcal{B}}_{R}$ through the area in parameter space enclosed by the pumping cycle.  The pseudo magnetic fields associated with charge and heat pumping through the double dot are shown in Figs.~\ref{fig_setup}(c)-(d). They exhibit features localized at the triple points.  If the pumping orbit is large enough to fully encircle one of these features, the charge and heat pumped per period become independent of the details of the pumping cycle and the resulting dc-currents show plateaux with height $\bar{I}^{(1)}_{\rm L}=\pm \frac{\Omega}{2\pi}$ and $\bar{Q}_{\rm L}^{(1)}=\pm \frac{\Omega}{2\pi}  T \ln 2$, see Fig.~\ref{fig:currents-omega1-eq}. 

While the appearance of plateaux in the charge current is directly related to the quantization of charge, plateaux in the heat current reflect  the degeneracies occurring in the double dot. This can be understood by noticing that along an orbit that fully encloses a triple point, whenever one of the two dots comes in resonance with its neighboring lead, the other one is strongly off-resonant, so that particles are exchanged only with one lead at the time. In this case, the average heat current flowing \HL{from} each lead $\bar{Q}_{r}^{(1)}$ is directly related to the
difference in entropy in the double dot before and after an electron has tunneled through the $r$ barrier, $\bar{Q}_{r}^{(1)}=T\Delta S^{(0)}_{r}$, where, $S^{(0)}=-\sum_{\xi}P_{\xi}^{(0)}\ln P_{\xi}^{(0)} $ is the Shannon entropy of the double dot and the subscript $r$ indicates that the 
entropy difference is between two charge configurations that differ only by one electron in dot $r$~\cite{Haupt13}. 
In this case where the double dot states are spin-degenerate, we have  $\Delta S^{(0)}_{r}=\pm\ln2$.

\subsection{Double-dot pump as a Carnot engine.} \label{sec:engine}
 An important feature of the double-dot pump  is the possibility of achieving quantized pumping even in the presence of a finite bias voltage $V$ or temperature gradient $\Delta T$. This requires minimizing the contributions of the instantaneous currents $\bar{R}^{(0)}$, which  play the role of leakage currents
\hl{since they flow in the direction set by the \HL{gradient,} irrespectively of the orientation of the pumping cycle. This} can be achieved by choosing a pumping cycle that fully encircles a single triple point. The triple point regions are broadened by $V$ and $\Delta T$. However, as long as they are well \HL{separated} from each other, it remains possible to pump one electron per cycle through the double dot 
and the heat currents in each lead \HL{exhibit} well defined plateaux of height $\frac{\Omega}{2\pi}T_{r}\ln 2 $, where $T_{r}$ is the local temperature of the lead~\cite{Haupt13}. In other words, even if the system is globally brought out of equilibrium, the heat exchanged \HL{solely with one} lead obeys Clausius relation $\bar{Q}_{r}^{(1)}=T_{r} \Delta S^{(0)}_{r}\frac{\Omega}{2\pi}$. This is because along an orbit that fully encircles a triple point, the double dot is effectively coupled only to one lead at the time. In the limit of slow driving $\Omega \to 0$, the system has time to equilibrate with the lead it is coupled to, so that processes that change the total occupation of the double dot represent isothermal transitions between equilibrium states.
A pumping cycle that fully encloses a triple point can then be regarded as a nanoscale analog of the Carnot cycle, in which two isothermal transitions are connected by two \hl{thermodynamically adiabatic transitions, i.e., occurring at constant entropy. The latter transitions} accompany the crossing of the two dots' levels.

When the double dot is operated between two leads with different temperatures and same chemical potential, it acts either as a refrigerator or as heat engine, depending on the direction of the cycle. In the first case, \HL{a power $\mathcal{P}_{\rm ac}$} has to be provided by the external ac-fields to extract heat from the cold reservoir. The efficiency of a refrigerator is \hl{characterized} by its coefficient of performance ${\rm COP}=\bar{Q}_{\rm cold}/\mathcal{P}_{\rm ac}$, where $\bar{Q}_{\rm cold}$ is the average heat-flow out of the cold reservoir. \emph{Vice versa}, if heat is extracted from the hot reservoir and released in the cold one, work is \hl{performed} on the external fields and the double dot functions as a heat engine. Its performance is characterized by the efficiency coefficient $\eta=(-\mathcal{P}_{\rm ac})/\bar{Q}_{\rm hot}$, with $(-\mathcal{P}_{\rm ac})$ the work per unit time {\em done by the system} on the ac-fields and $\bar{Q}_{\rm hot}$ the heat absorbed from the hot lead. 
It is  straightforward to show that in the ideal limit in which leakage currents can be completely neglected: we have ${\rm COP}=T_{\rm cold}/(T_{\rm hot}-T_{\rm cold})$ for the case of the cooling-cycle and $\eta=(T_{\rm hot}-T_{\rm cold})/T_{\rm hot}$ for the heat engine, meaning that the double-dot pump can be operated with Carnot efficiency for $\Omega\to0$~\cite{Haupt13}. 

The efficiency of a realistic double-dot engine, is however limited both by the leakage currents $\bar{R}^{(0)}$ and by dissipative effects associated to a finite driving frequency. These are captured  by the contribution to the currents to second order in the driving frequency, $\bar{R}^{(2)}$,
which for the heat current is of the order of $\bar{Q}_{r}^{(2)}\sim \delta_{E}\Omega^{2}/\Gamma$. This is the heat dissipated in each cycle due to the injection of hot electrons or holes into the leads.
As discussed in Ref.~\cite{Haupt13}, it can pose severe limitations to the efficiency of a double-dot based engine, especially in the regime 
of small $\Delta  T$. These limitations are less strict if one considers the operation of a double-dot pump as nanoscale ``battery charger'' performing work against a dc-source. In this case efficiencies up to 70\% of the ideal value can be achieved, see Ref.~\cite{Haupt13}.